# Generative Modeling in Sinogram Domain for Sparse-view CT Reconstruction

Bing Guan, Cailian Yang, Liu Zhang, Shanzhou Niu, Minghui Zhang, Yuhao Wang, *Senior Member, IEEE*, Weiwen Wu, *Member, IEEE*, Qiegen Liu, *Senior Member, IEEE*

*Abstract*—The radiation dose in computed tomography (CT) examinations is harmful for patients but can be significantly reduced by intuitively decreasing the number of projection views. Reducing projection views usually leads to severe aliasing artifacts in reconstructed images. Previous deep learning (DL) techniques with sparse-view data require sparse-view/full-view CT image pairs to train the network with supervised manners. When the number of projection view changes, the DL network should be retrained with updated sparse-view/full-view CT image pairs. To relieve this limitation, we present a fully unsupervised score-based generative model in sinogram domain for sparse-view CT reconstruction. Specifically, we first train a score-based generative model on full-view sinogram data and use multi-channel strategy to form high-dimensional tensor as the network input to capture their prior distribution. Then, at the inference stage, the stochastic differential equation (SDE) solver and data-consistency step were performed iteratively to achieve full-view projection. Filtered back-projection (FBP) algorithm was used to achieve the final image reconstruction. Qualitative and quantitative studies were implemented to evaluate the presented method with several CT data. Experimental results demonstrated that our method achieved comparable or better performance than the supervised learning counterparts.

*Index Terms*—Sparse-view CT, Sinogram domain, Generative modeling, Score-based diffusion.[1]

## I. INTRODUCTION

X-ray computed tomography (CT) has been widely applied owning to its remarkable ability of visualizing internal organs, bones, soft tissues and blood vessels [1]. However, it also brings the potential health risk with high radiation dose during clinical examinations. Reducing the projection views is an effective strategy to lower radiation dose. Unfortunately, the conventional filtered back projection (FBP) [2] algorithm will cause severe streak artifact with sparse-view sampling. The degraded image quality further imposes a big challenge for accurate clinical evaluation. Thus, achieving high-quality CT images from sparse-view data is an important topic.

Sparse-view CT reconstruction is a typically ill-posed inverse problem. Compressive sensing (CS) is a break-through in solving under-determined inverse systems especially for sparse-view CT reconstruction. For example, Sidky *et al*. [3] used the total variation (TV) as the regularization term to obtain promising results. Under the assumption of piece-wise constant, Yu and Wang [4] demonstrated the TV minimization compromised structural details and notorious blocky artifacts. Besides, other advanced priors such as nonlocal means [5], dictionary learning [6], low rank [7] were incorporated into iterative reconstruction. Furthermore, Zheng *et al*. [8] and Chun *et al*. [9] combined the penalized weighted-least squares with sparsity transforms for sparse-view CT reconstruction. However, for different CT imaging tasks, it is very difficult to find a universal regularization term to achieve consistently superior performance. Therefore, the deep learning-based sparse-view CT reconstruction achieved more attentions in recent years. These deep-learning-based methods mainly include image-based post-processing methods [10]-[20], projection domain interpolation methods [21]-[26], and dual-domain deep reconstruction methods [27]-[32].

The first class is the image-based post-processing methods. They directly process low-quality images as the input, which means that these methods do not need projection data [10-12]. This type is widely studied because they are not dependent on the imaging geometry of CT scanner [13], [14]. For example, the residual learning and U-net strategies were combined [15] to achieve promising denoising performance. Moreover, the wavelet transform was introduced into the Wave-Net [16] and Framing U-net [17]. These networks are updated with pixel-wise loss. Therefore, their results are susceptible to be over-smoothed. To address this issue, the perceptual loss [18] was introduced to generate more realistic CT images. These networks include generative adversarial network [19] and attribute augment image enhancement network [20]. These post-processing methods cannot effectively restore finer structures and features.

The second class is projection-domain interpolation methods, which directly estimate full-view data from sparse-view projections. For example, Lee *et al*. [21] employed a convolution neural network (CNN) [22] to interpolate the missing sinogram, where uses the residual learning for better convergence and the patch-wise training to avoid memory overload. Xu *et al*. [23] presented a residual learning-based U-Net architecture. Cho *et al*. [24] proposed a local linear interpolation method to synthesize missing data. Dong *et al*. [25] further introduced sinogram interpolation into iterative reconstruction framework. Lee *et al*. [26] trained a network to learn residual between input sinogram and sparsely sampled sinogram, and concatenated these two sinograms to generate full-view sinogram with an optimization network.

The dual-domain deep reconstruction methods concentrate to reconstruct high-quality images by considering the prior information in projection and image domains simultaneously.

This work was supported by National Natural Science Foundation of China (61871206, 62201628), and Science and Technology Program of Jiangxi Province (20192BCB23019, 20202BBE53024). (B. Guan and C. Yang are co-first authors.) (Corresponding authors: Weiwen Wu and Qiegen Liu.)
This work did not involve human subjects or animals in its research.

B. Guan, C. Yang, L. Zhang, M. Zhang, Y. Wang and Q. Liu are with School of Information Engineering, Nanchang University, Nanchang 330031, China. ({guanbing, yangcailian, zhangliu}@email.ncu.edu.cn, {zhangminghui, wangyuhao, liuqiegen}@ncu.edu.cn)
W. Wu is with the School of Biomedical Engineering, Sun Yat-Sen University, Shenzhen, Guangdong, China (wuweiw7@mail.sysu.edu.cn).
S. Niu is with the Ganzhou Key Laboratory of Computational Imaging, Gannan Normal University, Ganzhou 341000, China. (szniu@gnnu.edu.cn).

For example, Wurfl *et al.* [27] and Wang *et al.* [28] developed the back-projection layer to directly map projections to images. LEARN [29] unfolded the iterative framework into a deep-learning network, which is directly trained with projection data and the corresponding full-view images. Hu *et al.* [30] proposed a hybrid domain neural network (HDNet), which recovered projection and image information successively. Liu *et al.* [31] presented a novel cascade model LS-AAE that mainly focuses on availably utilizing the spatial information between greatly correlated images for dual domain sparse-view reconstruction. Zhang *et al.* [32] designed a hybrid-domain convolutional neural network for limited-angle CT. Wu *et al.* [13] presented a dual-domain residual-based optimization network (DRONE), which performed well in edge preservation and detail recovery. However, these methods may suffer from secondary artifacts in the reconstructed images caused by projection interpolation.

In this study, we propose an unsupervised DL modeling in sinogram domain for sparse-view CT reconstruction, namely GMSD. Specifically, we use a score-based generative model to generate the full-view sinogram from sparse-view projection data. Our main idea is to learn the prior distribution of full-view projection with a generative model, which is explored to infer the missing data with sparse-view sampling. Unlike previous supervised methods, our unsupervised method does not require to retrain on sparse-view/full-view reconstruction pairs with changed projection views.

In fact, the image features from data domain are different from the image-domain features. On the one hand, the measurement data within Radon domain has the shape of sine, this property will be corrupted with angle information missing in the case of sparse-view data reconstruction. The trained neural network would explore the natural property within the missing views by training projection data. In the other hand, different neighbor detector units from the same view also share similar data. The trained neural network in the projection data will help to further explore the implicit relationship from detector element. More importantly, the data consistency module is necessary to most of the image-based score matching model. It is difficult to compute the projection and backprojection operations in circular/spiral cone-beam CT. However, such operations can be avoided in the projection generative models. Thus, we developed our GMSD reconstruction method in the projection domain rather than image-domain. The main contributions of this work are summarized as follows:

- Designing a score-based generative model for projection domain interpolation. First, multiple scales noise are used to perturb the data. Then, the data fidelity and prior terms are alternative updated after the prediction and correction steps to ensure optimal solutions.
- With the unsupervised training mechanism, there is no need to update the training datasets and retrain the model when the projection views changed.

The rest of the manuscript is organized as follows. Relevant background on score-based diffusion models is described in Section II. Detailed procedure and algorithm is presented in Section III. Experimental results and specifications about the implementation are given in Section IV. We conclude our work in Section V.

## II. PRELIMINARY

### A. Deep Generative Models

Recently advances with deep generative networks have shown obvious gains in modeling complex distributions such as images [33], audios [34] and texts [35]. The popular deep generative models can be primarily categorized into two groups: Explicit generative model and implicit generative model. The former model provides an explicit parametric specification of the data distribution, including autoencoders (AE) and its variants [36], [37], flow-based generative models [38], [39], score-based models [43] and deep Boltzmann machine [40]. Specially, for a log-likelihood function $\log p(x)$, score-based models train parametric network to approximate the likelihood gradient $\nabla_x \log p(x)$. Alternatively, we can specify implicit probabilistic models that define a stochastic procedure to directly generate data. GANs [41] are the well-known implicit likelihood models. They optimize the objective function using adversarial learning and have been shown to produce high quality images [42]. Due to the occurrence of pattern collapse, GANs still suffer from a remarkable difficulty in training.

### B. Score-based SDE

Score-based diffusion models perturb the data distribution according to the forward SDE by injecting Gaussian noise, arriving at a tractable distribution (e.g., isotropic Gaussian distribution). To sample the data from this distribution, one can train a neural network to estimate the gradient of the log data distribution (i.e., score function $\nabla_x \log p(x)$), which can be used to solve the reverse SDE numerically. The key idea is summarized in Fig. 1.

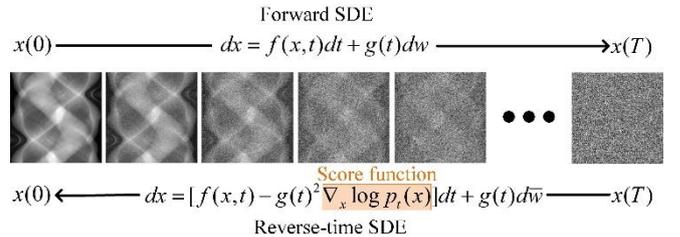

**Fig. 1.** The perturbed data by noise is smoothed with following the trajectory of an SDE. By estimating the score function $\nabla_x \log p_t(x)$ with SDE, it is possible to approximate the reverse SDE and then solve it to generate image samples from noise.

$\{x(t)\}_{t=0}^T$ with $x(t) \in \mathbb{R}^n$, where $t \in [0,T]$ is the time index of the progression and $n$ denotes the sinogram image dimension. We choose $x(0) \sim p_0$ and $x(T) \sim p_T$, $p_0$ is the data distribution and $p_T$ is the prior distribution. Then, the diffusion process can be modeled as the solution to the following SDE:

$$dx = f(x,t)dt + g(t)dw \quad (1)$$

where $f(x,t) \in \mathbb{R}^n$ and $g(t) \in \mathbb{R}$ correspond to the drift and diffusion coefficients, respectively. $w \in \mathbb{R}^n$ induces the Brownian motion.

One can construct different SDEs by choosing different functions $f(x,t)$ and $g(t)$. Since variance exploding (VE) SDEs can lead to higher sample qualities, here we develop the following VE-SDE:

$$f(x,t) = 0, \quad g(t) = \sqrt{\frac{d[\sigma^2(t)]}{dt}} \quad (2)$$

where $\sigma(t) > 0$ is a monotonically increasing function, which is typically chosen to be a geometric series [46], [47].

Starting from samples of $x(T) \sim p_T$ and reversing the process, we can obtain samples $x(0) \sim p_0$. It is well known that the reverse of a diffusion process is also a diffusion process. Thus, we obtain the following reverse-time SDE:

$$dx = [f(x,t) - g(t)^2 \nabla_x \log p_t(x)]dt + g(t)d\overline{w}$$
$$= \frac{d[\sigma^2(t)]}{dt}\nabla_x \log p_t(x) + \sqrt{\frac{d[\sigma^2(t)]}{dt}}d\overline{w} \qquad (3)$$

where $\overline{w}$ is a standard Brownian motion with time flows backwards from $T$ to $0$, and $dt$ is an infinitesimal negative timestep. Once the score of each marginal distribution $\nabla_x \log p_t(x)$ is known for all $t$, we can derive the reverse diffusion process from Eq. (1) and simulate it to sample from $p_0$.

In order to solve Eq. (3), one has to know the score function for all $t$, which can be estimated using a time-conditional neural network $S_\theta(x,t) \simeq \nabla_x \log p_t(x(t))$. The training parameters $\theta$ includes all the weights and bias raised in interpolation and denoising networks. Since we do not know the true score, we can use denoising score matching [46] to replace the unknown $\nabla_x \log p_t(x)$ with $\nabla_x \log p_t(x(t)|x(0))$, where $\nabla_x \log p_t(x(t)|x(0))$ is the Gaussian perturbation kernel centered at $x(0)$. Under some regularity conditions, $S_\theta(x,t)$ trained with denoising score matching will satisfy $S_{\theta^*}(x,t) = \nabla_x \log p_t(x(t))$ almost surely [43]. Formally, we optimize the parameters $\theta$ of the score network with the following cost:

$$\theta^* = \arg\min_\theta \mathbb{E}_t\{\lambda(t)\mathbb{E}_{x(0)}\mathbb{E}_{x(t)|x(0)}[\\ \|S_\theta(x(t),t) - \nabla_{x(t)} \log p_t(x(t)|x(0))\|^2]\} \qquad (4)$$

Once the network is trained with Eq. (4), we can plug the approximation $S_\theta(x,t) \simeq \nabla_x \log p_t(x(t))$ to solve the reverse SDE in Eq. (3):

$$dx = \frac{d[\sigma^2(t)]}{dt}S_\theta(x,t) + \sqrt{\frac{d[\sigma^2(t)]}{dt}}d\overline{w} \qquad (5)$$

Then, we can solve the SDE numerically with Euler discretization [44]. This involves discretizing $t$ in range $[0,1]$, which is uniformly separated into $N$ intervals such that $0 = t_0 < \cdots < t_N = 1$, $\Delta t = 1/N$. Additionally, we can correct the direction of gradient ascent with Langevin Markov Chain Monte Carlo algorithm [47]. Iteratively implementing predictor and corrector step yields the predictor-corrector sampling algorithm [44] that presented in **Algorithm 1.**

| Algorithm 1: Predictor-Corrector (PC) sampling |
|---|
| **Setting:** $S_\theta, N, M, \sigma, \varepsilon$ |
| 1: $x^N \sim \mathbb{N}(0, \sigma_{max}^2)$ |
| 2: **For** $i = N-1$ to $0$ **do** |
| 3: $\quad \tilde{x}^i \leftarrow x^{i+1} + (\sigma_{i+1}^2 - \sigma_i^2)S_\theta(x^{i+1}, \sigma_{i+1})$ |
| 4: $\quad z \sim \mathbb{N}(0,1)$ |
| 5: $\quad x^i \leftarrow \tilde{x}^i + \sqrt{\sigma_{i+1}^2 - \sigma_i^2} z_{i+1}$     **Predictor** |
| 6: $\quad$ **For** $j = 1$ to $M$ **do** |
| 7: $\quad\quad z \sim \mathbb{N}(0,1)$ |
| 8: $\quad\quad \tilde{x}^{i,j-1} \leftarrow x^{i,j-1} + \varepsilon_i S_\theta(x^{i,j-1}, \sigma_i)$ |
| 9: $\quad\quad x^{i,j} \leftarrow \tilde{x}^{i,j-1} + \sqrt{2\varepsilon_i} z_i$     **Corrector** |
| 10: $\quad$ **End for** |
| 11: **End for** |
| 12: **Return** $x^0$ |

## III. PROPOSED GMSD

### A. GMSD Imaging Model

In this study, the linear measurement process for sparse-view CT imaging is visualized in Fig. 2. Intuitively, $diag(\Lambda)$ can be viewed as a subsampling mask on the sinogram, and $P(\Lambda)$ subsamples the sinogram into an observation $y$ with a smaller size according to this subsampling mask. $I$ represents CT images. $T$ corresponds to the Radon transform. The CT reconstruction problem can be formulated as finding sinogram image $x$ given the following equation:

$$y = P(\Lambda)TI = P(\Lambda)x \qquad (6)$$

where $x$ denotes the full-view sinogram data, $y$ is the sparse-view CT projection data.

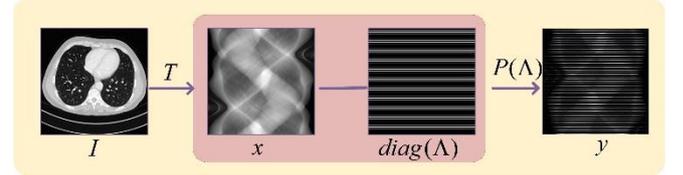

**Fig. 2.** Linear measurement processes for sparse-view CT.

In the circumstance of sparse-view projection, it is easy to produce strip artifacts if the reconstructed image is directly obtained by FBP algorithm. In order to obtain higher quality reconstructed images, we need to obtain full-view projection data from sparse-view projection data, which itself is an underdetermined inverse problem. To solve this problem, various priors are incorporated into regularized objective function. The regularized objective function is expressed as:

$$x = \arg\min_x \|P(\Lambda)x - y\|_2^2 + \tau R(x) \qquad (7)$$

where the first term is data fidelity, which determines the consistency between the desired full-view data $x$ and observed measurement $y$. The second term $R(x)$ is regularization term, and $\tau$ is used to balance data fidelity and regularization.

Inspired by [48]-[50], we introduce a high-dimensional space embedding strategy to boost the representation diversity of generative modeling, and thereafter reconstruction performance. The underlying supporting theoretical can be found in [48]. In detail, we utilize a channel-copy transformation $X = H(x)$ to establish a $N$-channel higher-dimensional tensor, e.g., the vector variable with $N$ channels as $X = [x,x,\cdots,x]$. The goal of stacking $X$ is to form object in high-dimensional manifold, which can improve accuracy of score estimation. Subsequently, the GMSD is trained with $X$ in high-dimensional space as network input and the parameterized $S_\theta(X,t)$ is obtained. Figure 3 (a) illustrates the multi-channel strategy in the image domain of our previous work [50], and Fig. 3 (b) shows the multi-channel strategy in projection domain used in this work. Mathematically, Eq. (7) is further expressed as:

$$x = \arg\min_x \|P(\Lambda)x - y\|_2^2 + \tau R(X) \qquad (8)$$

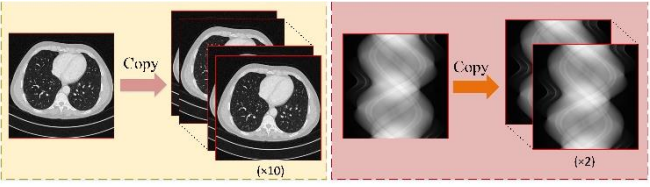

**Fig. 3.** Multi-channel strategy in different methods. (a) EASEL (b) GMSD.

*B. Outer-loop of GMSD*

To obtain richer information and high-quality results, the prior distribution of sinogram data is estimated using the score-based generative models. Specifically, instead of perturbing data with a finite number of noise distributions, a continuous distribution over time is considered with a diffusion process. By reversing SDE, we can convert random noise into data for sampling. For samples update step, as suggested by [44], the predictor-corrector (PC) sampling is used in this work. Langevin dynamics is considered as the corrector, which transforms any initial sample $X(0)$ to an approximate sample from $p_t(X)$ with the following procedure:

$$X^{i,j} \leftarrow X^{i,j-1} + \varepsilon_i S_\theta(X^{i,j-1}; \sigma_i) + \sqrt{2\varepsilon_i} z_i \quad (9)$$
$$j = 1, 2, \cdots, M, \quad i = N-1, \cdots, 0$$

where $\varepsilon_i > 0$ is the step size, and $z_i \sim N(0,1)$ is standard normalization. The above is repeated for $i = N-1, \cdots, 0$. The theory of Langevin dynamics guarantees that when $j \to \infty$ and $\varepsilon_i \to 0$, $X^i$ is a sample from $p_t(X)$ under some regularity conditions.

The above-mentioned sampling is not directly from $p(X)$, but from the posterior distribution $p(X|y)$ by employing SDE in Section II. An intuitive solution is that the data-consistency in Eq. (6) can be considered as a conditional term which can be incorporated into the sampling procedure of Eq. (9). i.e., let $x^{i,j} = Mean(X^{i,j})$, it yields:

$$\begin{aligned} x^{i,j} &= x^{i,j-1} + \varepsilon_i \nabla_X \left[ \log p_t(X^{i,j-1}) + \lambda \log p(y|x^{i,j-1}) \right] + \sqrt{2\varepsilon_i} z_i \\ &= x^{i,j-1} + \varepsilon_i S_\theta(X^{i,j-1}; \sigma_i) + \varepsilon_i \lambda \nabla_x \left\| y - P(\Lambda) x^{i,j-1} \right\|^2 + \sqrt{2\varepsilon_i} z_i \end{aligned}$$
(10)

where $\log p(y|x)$ is given by the data model that derived from data knowledge. The $\log(p(X))$ is given by the prior model that represents information known beforehand about the true model parameter. The hyperparameter $\lambda$ balances the trade-off between priors and data fidelity.

The predictor refers to a numerical solver for the reverse-time SDE. Specifically, the samples from the prior distribution can be obtained from the reverse SDE in Eq. (3), which can be discretized as follows:

$$X^i = X^{i+1} + (\sigma_{i+1}^2 - \sigma_i^2) S_\theta(X^{i+1}, \sigma_{i+1}) + \sqrt{\sigma_{i+1}^2 - \sigma_i^2} z_{i+1} \quad (11)$$

where $X(0) \sim p_0$, and we set $\sigma_0 = 0$ to simplify the notation. Then, adding the data-consistency item constraint to Eq. (11), we obtain the following formula:

$$\begin{aligned} x^i &= x^{i+1} + (\sigma_{i+1}^2 - \sigma_i^2) S_\theta(X^{i+1}, \sigma_{i+1}) + \sqrt{\sigma_{i+1}^2 - \sigma_i^2} z_{i+1} \\ &\quad + \lambda \nabla_x \left\| y - P(\Lambda) x^{i+1} \right\|^2 \end{aligned}$$
(12)

*C. Inner-loop of GMSD*

Here, the details for implementing Eq. (10) and Eq. (12) are further given. We first use an alternating optimization algorithm to minimize the decoupling of prior information items and data-consistency items. Then, two sub-problems are alternately update to achieve the optimal solution. After the virtual variable $U^{i,j}$ is updated, we turn it back to the original variable $u^{i,j} = Mean(U^{i,j})$ via channel-average operator. Especially, Eq. (10) is decomposed into two steps as follows:

$$U^{i,j} = X^{i,j-1} + \varepsilon_i S_\theta(X^{i,j-1}; \sigma_i) + \sqrt{2\varepsilon_i} z_i \quad (13)$$

$$x^{i,j} = \arg\min_x [\| y - P(\Lambda) x^{i,j-1} \|^2 + \beta \| x^{i,j-1} - u^{i,j} \|^2] \quad (14)$$

where $\beta = \varepsilon_i \lambda$, hyperparameter $\lambda = 1$, $i = N-1, \cdots, 0$ and $j = 1, 2, \cdots, M$ represents the iteration of outer and inner loops. Specifically, the second minimization in Eq. (14) is a standard least square (LS), which can be solved as follows:

$$x^{i,j} = \frac{P(\Lambda)^T (P(\Lambda) x^{i,j-1} - y) + \beta (x^{i,j-1} - u^{i,j})}{P(\Lambda)^T P(\Lambda) + \beta} \quad (15)$$

Meanwhile, Eq. (12) can be decoupled into the following two steps:

$$V^i = X^{i+1} + (\sigma_{i+1}^2 - \sigma_i^2) S_\theta(X^{i+1}, \sigma_{i+1}) + \sqrt{\sigma_{i+1}^2 - \sigma_i^2} z_{i+1} \quad (16)$$

$$x^i = \arg\min_x [\| y - P(\Lambda) x^{i+1} \|^2 + \beta \| x^{i+1} - v^i \|^2] \quad (17)$$

Eq. (17) can be further reduced as:

$$x^i = \frac{P(\Lambda)^T (P(\Lambda) x^{i+1} - y) + \beta (x^{i+1} - v^i)}{P(\Lambda)^T P(\Lambda) + \beta} \quad (18)$$

Once we obtain the full-view projection $x$, the final image is obtained:

$$\tilde{I} = FBP(x) \quad (19)$$

In summary, the flowchart of training phase and iterative reconstruction phase for GMSD is shown in Fig. 4. Specifically, we first train a score-based generative model to capture projections prior distribution. Then, at the inference stage, the numerical SDE solver and data-consistency step is iteratively updated to achieve the full-view projection. Since the Langevin dynamics updating can guarantee algorithms convergence, the overall GMSD algorithm will be convergent after finite iterations.

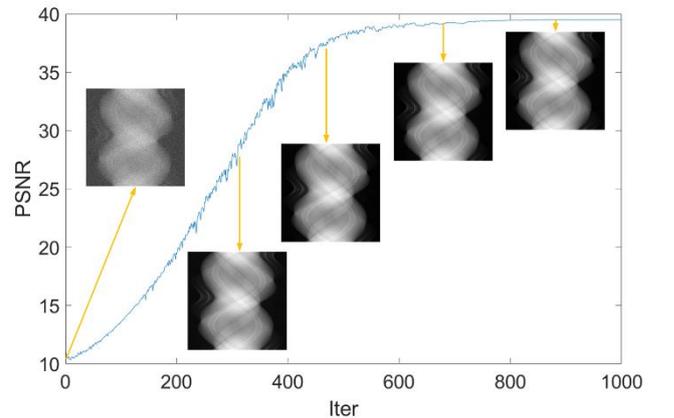

**Fig. 5.** Visualization of the intermediate reconstruction process of GMSD. As the level of artificial noise becomes smaller, the reconstruction results tend to ground-truth data.

To validate the convergence of GMSD, the intermediate samples associated with PSNR values of reconstructing CT images from 120 views are given in Fig. 5, where samples evolve from pure random noise to reconstructed images. The fluctuation of the curve is not obvious as the iteration increases. In addition, the whole reconstruction process is convergent and basically stable at 500 iterations. Therefore, GMSD has a reasonably fast convergence rate.

Furthermore, **Algorithm 2** describes the reconstruction algorithm in detail. The whole GMSD reconstruction process includes two loops. The outer loop is the entire PC sampling process. Prediction is first performed with trained network and then the inner loop is performed for correction. The data prior items and data fidelity items of the prediction and correction process are updated in each loop.

| Algorithm 2: GMSD for iterative reconstruction |
|---|
| **Training stage** |
| **Dataset:** Multi-channel dataset: $X = [x, x]$ |
| 1: Choose the network architecture |
| 2: Trained DMSP $S_\theta(X, t)$ |
| **Reconstruction stage** |
| **Setting:** $S_\theta, N, M, \sigma, \varepsilon$ |
| 1: $X^N \sim \mathbb{N}(0, \sigma_{\max}^2)$ |
| 2: **For** $i = N-1$ to $0$ **do (Outer loop)** |
| 3: $\quad V^i \leftarrow Predictor(V^{i+1}, \sigma_i, \sigma_{i+1})$ |
| 4: $\quad v^i = Mean(V^{i,j})$ |
| 5: $\quad Update\ x^{i,j}$ by Eq. (15) **(Data-consistency)** |
| 6: $\quad$ **For** $j = 1$ to $M$ **do** |
| 7: $\quad\quad U^{i,j} \leftarrow Corrector(U^{i,j-1}, \sigma_i, \varepsilon_i)$ **(Inner loop)** |
| 8: $\quad\quad u^{i,j} = Mean(U^{i,j})$ |
| 9: $\quad\quad Update\ x^i$ by Eq. (18) **(Data-consistency)** |
| 10: $\quad$ **End for** |
| 11: **End for** |
| 12: Final image $\tilde{I} = FBP(x)$ |
| 13: Return $\tilde{I}$ |

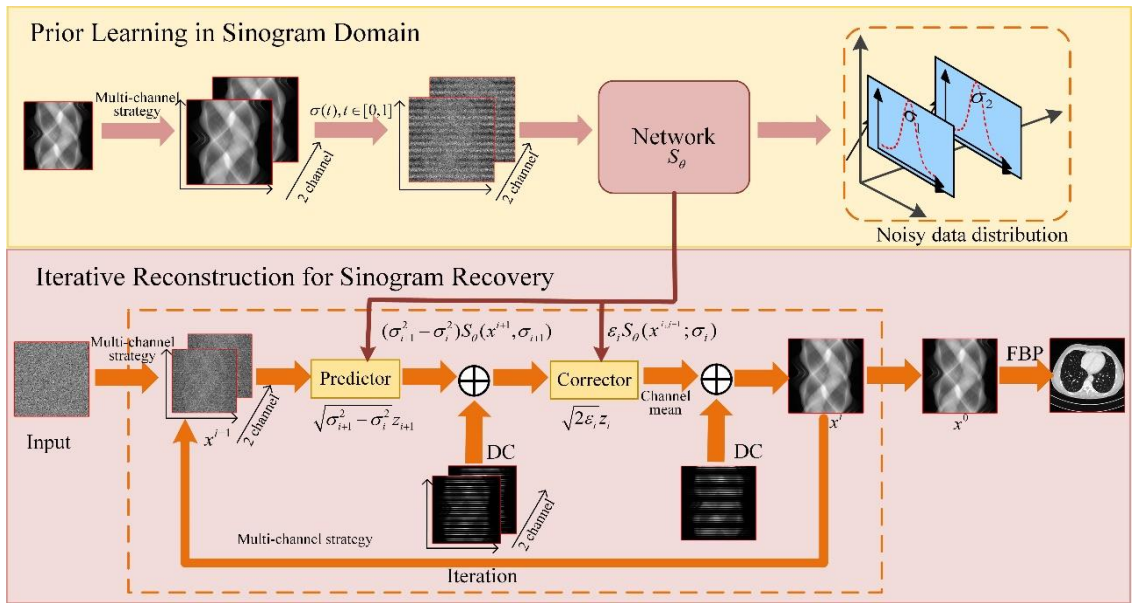

**Fig. 4.** The proposed unsupervised deep learning in sinogram domain for sparse-view CT. Top: Training stage to learn the gradient distribution via denoising score matching. Bottom: Iterate between numerical SDE solver and data-consistency step to achieve reconstruction. DC stands for the data-consistency items.

## IV. EXPERIMENTS

### A. Data Specification

**AAPM Challenge Data:** The simulated data from human abdomen images provided by Mayo Clinic for the AAPM Low Dose CT Grand Challenge [51] are used for evaluation. The data includes high-dose CT scans from 10 patients, where 9 of them for training and the other one for evaluation. For parallel-beam reconstruction network training, 5388 slices with 1 $mm$ thickness and each of them covering $512 \times 512$ pixels are employed, where 4839 and 549 images are used to training and testing. Artifact-free images are generated from 720 projection views by FBP algorithm which can be considered as reference. We extracted 60-, 90-, 120-, and 180-view projection data from full-view projection for sparse-view CT reconstruction. In fan-beam CT reconstruction, Siddon's ray-driven algorithm [52], [53] is utilized to generate the projection data. The distance from rotation center to the source and detector are set to 40 $cm$ respectively. The detector width is 41.3 $cm$ including 720 detector elements and a total of projection views are evenly distributed over $360°$.

**CIRS Phantom Data:** A high-quality set of CT volumes ($512 \times 512 \times 100$ voxels, voxel size $0.78 \times 0.78 \times 0.625$ $mm^3$) of an anthropomorphic CIRS phantom is obtained from a GE Discovery HD750 CT system, in which the tube current value is set to 600 mAs. The source-to-axial distance is 573 mm, and the source-to-detector distance is 1010 mm. We extracted 60-, 90-, 120-, and 180-view projection data for sparse-view CT reconstruction.

### B. Model Training and Parameter Selection

In our experiments, the model is trained by the Adam algorithm with the learning rate $10^{-3}$ and Kaiming initialization is used to initialize the weights. The method is implemented in Python using Operator Discretization Library (ODL) [54] and PyTorch on a personal workstation with a GPU card (Tesla V100-PCIE-16GB). In the reconstruction stage, the iteration number is set to $N = 1000$, $M = 2$. Each time the prediction process of the outer loop is executed, the correction process of the inner loop is iterated twice by annealing Langevin. Our source code is publicly access at: https://github.com/yqx7150/GMSD.

### C. Quantitative Indices

To evaluate the quality of the reconstructed data, peak signal-to-noise ratio (PSNR), structural similarity index (SSIM), and mean squared error (MSE) are used for quantitative assessment.

PSNR describes the relationship of the maximum possible power of a signal with the power of noise corruption. Higher PSNR means better reconstruction quality. Denoting $I$ and $\tilde{I}$ to be the estimated reconstruction and ground-truth, PSNR is expressed as:

$$PSNR(I,\tilde{I}) = 20\log_{10}\left[\text{Max}(\tilde{I})/\left\|I-\tilde{I}\right\|_2\right] \quad (20)$$

The SSIM value is used to measure the similarity between the ground-truth and reconstruction. SSIM is defined as:

$$SSIM(I,\tilde{I}) = \frac{(2\mu_I\mu_{\tilde{I}}+c_1)(2\sigma_{I\tilde{I}}+c_2)}{(\mu_I^2+\mu_{\tilde{I}}^2+c_1)(\sigma_I^2+\sigma_{\tilde{I}}^2+c_2)} \quad (21)$$

where $\mu_I$ and $\sigma_I^2$ are the average and variances of $I$. $\sigma_{I\tilde{I}}$ is the covariance of $I$ and $\tilde{I}$. $c_1$ and $c_2$ are used to maintain a stable constant.

MSE is a measure of errors between paired observations expressing the same phenomenon. It is defined as:

$$MSE(I,\tilde{I}) = \sum_{i=1}^{W}\left\|I_i-\tilde{I}_i\right\|_2/W \quad (22)$$

where $W$ is the number of pixels within the reconstruction result. If MSE approaches to zero, the reconstructed image is closer to the reference image.

### D. Parallel-beam CT Reconstruction

*AAPM Challenge Data Study:* We compare our proposed method with four baseline techniques in sparse-view CT reconstruction including FBP [2], FISTA-TV [55], and SART-TV [56]. The involved parameters are set following the guidelines in their original papers.

For sparse-view CT reconstruction with 60, 90, 120 and 180 projection views, the PSNR and SSIM values of the reconstructed results from AAPM Challenge Dataset are listed in Table 1. The best PSNR and SSIM values of the reconstructed images with different projection views are highlighted in bold. It is obvious that GMSD presents more gains compared to the other methods. It can be observed that compared with other three methods, GMSD is able to achieve impressive average PSNR gain of 9.59 dB, 8.19 dB, 8.27 dB and 8.66 dB at 60, 90, 120, 180 views in Table 1. The reconstructed images contain less artifacts and more details with increased projection views. It is exciting that the image reconstructed by GMSD method can reach 35.95 dB in the case of 60 projection views. Meanwhile, the results reconstructed by GMSD method have the highest SSIM values compared to other competing algorithms. Thus, GMSD can achieve visible gains in terms of noise and artifacts suppression.

To further illustrate the merits of GMSD method, the reconstructed images and residual images are depicted in Figs. 6-7. As shown in Fig. 6, the FBP method results are the worst, as the performance of the analytical approach highly relies on the number of projection views. The results of FISTA-TV outperform those obtained by FBP algorithm. The sharper result with reasonable boundaries is obtained from the deep prior reconstruction approach. However, it still suffers from some undesirable artifacts and detail missing. Although SART-TV algorithm achieves acceptable result, some edge details are still lost. On the contrary, the image reconstructed by GMSD method is very close to the ground truth with preserved details and structures.

TABLE I
RECONSTRUCTION PSNR/SSIM/MSEs OF AAPM CHALLENGE DATA USING DIFFERENT METHODS AT 60, 90, 120 AND 180 VIEWS.

| Views | FBP | FISTA-TV | SART-TV | GMSD |
|---|---|---|---|---|
| 60 | 23.08/0.5936/0.00502 | 24.73/0.8532/0.00380 | 31.27/0.9321/0.00097 | **35.95/0.9662/0.00027** |
| 90 | 26.29/0.7053/0.00241 | 27.26/0.9199/0.00217 | 36.26/0.9583/0.00028 | **38.13/0.9763/0.00017** |
| 120 | 28.39/0.7920/0.00148 | 28.04/0.9458/0.00180 | 38.56/0.9666/0.00015 | **39.94/0.9828/0.00011** |
| 180 | 32.35/0.8965/0.00060 | 28.59/0.9577/0.00155 | 39.72/0.9700/0.00011 | **42.22/0.9881/0.00007** |

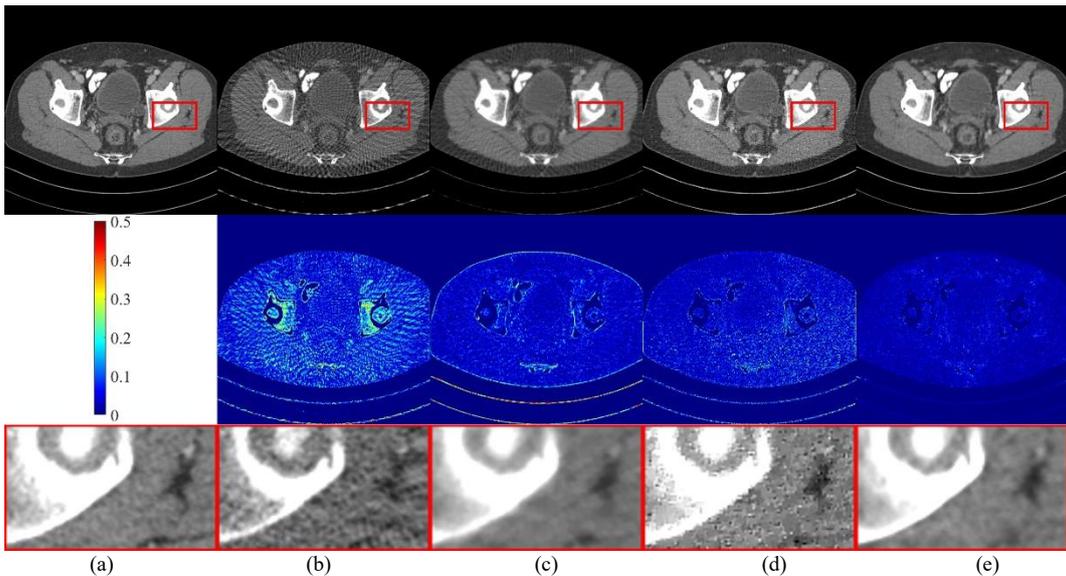

**Fig. 6.** Reconstruction results from 120 views using different methods. (a) The reference image versus the images reconstructed by (b) FBP, (c) FISTA-TV, (d) SART-TV, and (e) GMSD. The display windows are [-240,360]. The second row is residuals between the reference images and reconstruction images.

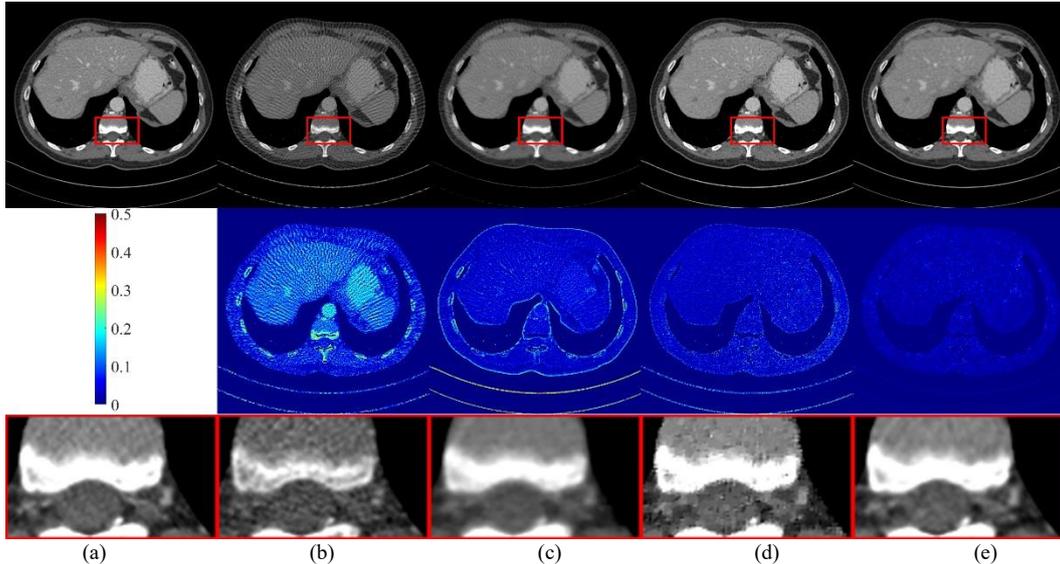

**Fig. 7.** Reconstruction images from 180 views using different methods. (a) The reference image versus the images reconstructed by (b) FBP, (c) FISTA-TV, (d) SART-TV, and (e) GMSD. The display windows are [-200,400]. The second row is residuals between the reference images and reconstruction images.

Compared with the results reconstructed from 120 projection views, the results reconstructed from 180 projection views exhibit obvious improvements of image quality as indicated in Figs. 6-7. The image reconstructed by FBP algorithm not only contains artifacts, but also some important structural features are not well preserved. Meanwhile, images reconstructed by FISTA-TV still suffer from severe steaking artifacts. Moreover, SART-TV can distinguish some details, but the edges are over-smoothed. Finally, the image reconstructed by GMSD method preserves more structural details while suppressing streaking artifacts.

*E. Fan-beam CT Reconstruction*

AAPM Challenge Data Study: For fan-beam CT reconstruction, in addition to the FBP [2], two supervised deep reconstruction methods including U-Net [57] and FBPConvNet [58] are added for evaluation. [57] implemented a U-Net structure for inter-polating sparsely sampled singoram to reconstruct CT images by an FBP algorithm. We calculate the average value of the test set and present the results in Table 2 to verify the effectiveness of GMSD. Table 2 lists the quantitative results for the reconstructions from 60, 90, 120, and 180 views, where the best results are highlighted in bold. It can be observed that GMSD method yields the second best results in terms of PSNR and SSIM values, which are consistent with the visual effects. Specifically, compared to FBPConvNet method, GMSD method has a notable gain of 0.46 dB, 1.92 dB, 1.6 dB and 0.96 dB in cases of 60, 90, 120 and 180 views. It is worth noting that our results are also impressive in terms of SSIM. From Table II, we can easily find out that DRONE surpasses the others in most aspects, except for PSNR in 180 views. However, DRONE is a complicated model with dual-domain and, compared with GMSD, it shows low generalization ability.

In Fig. 8(b), as the projection view reduces to 60, streak artifacts gradually increase in the image reconstructed by FBP algorithm. The quality of FBP image is so poor that some details and important structures cannot be distinguished. U-Net method can remove some artifacts while some important details are lost. The edges of the results reconstructed by FBP-ConvNet method is blurred as indicated in Fig. 8(d). On the contrary, GMSD results in Fig. 8(f) can achieve the second best performance in terms of noise suppression and structural detail preservation.

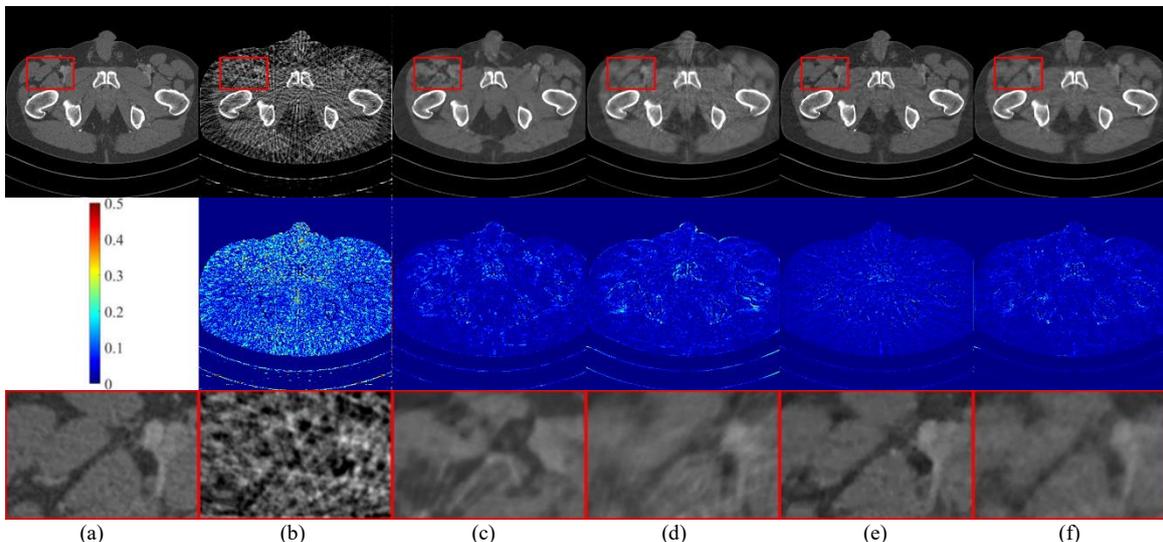

**Fig. 8.** Reconstruction images from 60 views using different methods. (a) The reference image versus the images reconstructed by (b) FBP, (c) U-Net, (d) FBPConvNet, (e) DRONE, and (f) GMSD. The display windows are [-250,600]. The second row is residuals between the reference images and reconstruction images.

TABLE II
RECONSTRUCTION PSNR/SSIM/MSEs OF AAPM CHALLENGE DATA USING DIFFERENT METHODS AT 60, 90, 120 AND 180 VIEWS.

| Views | FBP | U-Net | FBPConvNet | DRONE | GMSD |
|---|---|---|---|---|---|
| 60 | 21.15/0.4857/0.00785 | 27.48/0.9070/0.00214 | 35.64/0.9658/0.00029 | **36.93/0.9731/0.00027** | 36.10/0.9700/0.00028 |
| 90 | 23.60/0.6057/0.00450 | 32.36/0.9528/0.00067 | 37.09/0.9754/0.00022 | **39.91/0.9837/0.00012** | 39.01/0.9803/0.00014 |
| 120 | 26.30/0.7077/0.00241 | 36.12/0.9743/0.00026 | 39.47/0.9826/0.00012 | **41.26/0.9874/0.00008** | 41.07/0.9854/**0.00008** |
| 180 | 30.19/0.8475/0.00099 | 38.77/0.9799/0.00015 | 42.27/0.9884/0.00006 | 42.87/**0.9914/0.00005** | **43.23**/0.9909/**0.00005** |

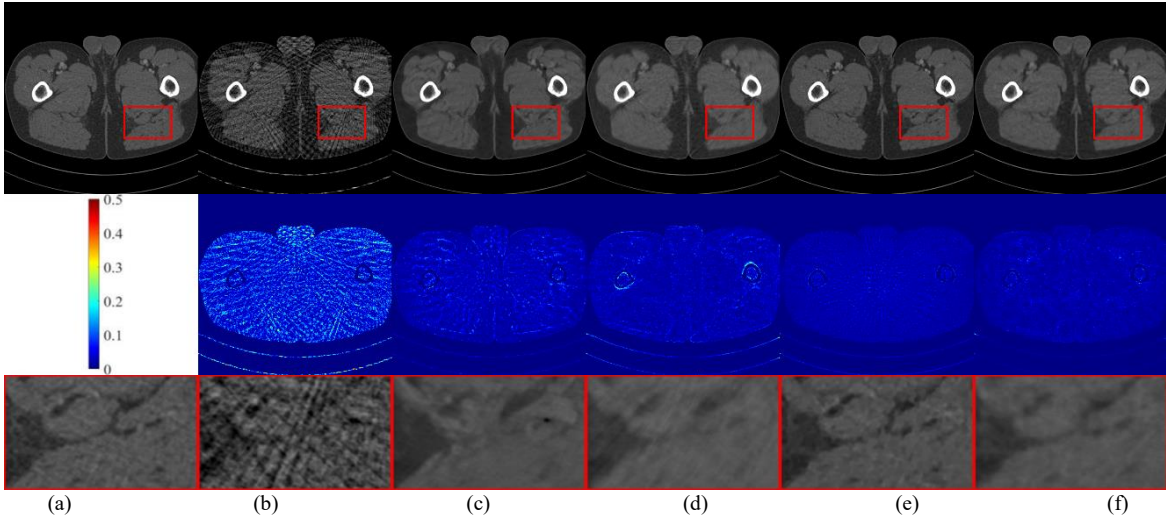

**Fig. 9.** Reconstruction images from 90 views using different methods. (a) The reference image versus the images reconstructed by (b) FBP, (c) U-Net, (d) FBPConvNet, (e) DRONE, and (f) GMSD. The display windows are [-250,600]. The second row is residuals between the reference images and reconstruction images.

Fig. 9 depicts the reconstructed images from 90 views with different methods. Compared with the images in Fig. 8(b), the artifacts in the image reconstructed by FBP algorithm are reduced as shown in Fig. 9(b). In Fig. 9(c), the U-Net compromises structural details and suffers from notorious blocky artifacts. Some details in FBPConvNet are over-smoothed. In Fig. 9(f), GMSD can retain more structure information and details than other method.

*CIRS Phantom Data Study:* To further validate the robustness of the proposed unsupervised learning scheme, we learn the prior knowledge on AAPM challenge data and test our model on CIRS phantom data. Table III records all the results, and the best value of each metric is marked in black bold. Intuitively, our method scores the highest PSNR. At the same time, FBP, U-Net and FBPConvNet methods obtain poor image quality evaluation metrics as compared with GMSD method. It implies that the comparison algorithms still lose some details and suffer from remaining artifacts. Compared with FBPConvNet, U-Net has more advantages on CIRS dataset. Because FBPConvnet is a sparse-view reconstruction method in image domain, which learns the feature information of the image, while U-Net is a reconstruction algorithm in projection domain, which learns the projection mapping relationship. Meanwhile, the reconstructed results of GMSD are 3.96 dB, 5.15 dB, 5.46 dB and 5.43 dB higher than U-Net at the case of 60, 90, 120, and 180 views, respectively. Therefore, it implies that GMSD has great adaptability on different datasets.

To visually illustrate the performance of GMSD, we perform qualitative comparisons of different views. In addition, for better evaluation of image quality, Fig. 10 and Fig. 11 depict the zoomed regions-of-interest (ROI) as marked by the red rectangles. From the results in Fig. 10(b), we can see that FBP reconstruction leads to severely degraded CT images with obviously noise and artifacts. Furthermore, it also can be seen that FBPConvNet effectively reduce noise. However, the details and texture information are lost. Comparing with the results in Fig. 10(b)-(d), we can see that GMSD method achieves the best performance in terms of noise-artifact suppression and tissue feature preservation. Meanwhile, U-Net performs well but tends to over-smoothed textures and edges as indicated by Fig. 11. Compared with the competitive reconstruction methods, GMSD performs better in terms of noise-artifact reduction and detail preservation.

### F. Preclinical Mouse Validation

The generalization is a major issue for deep learning-based CT reconstruction. We are often confronted with such situations: (1) The training and test datasets come from different machines, and (2) the noise level within the training and test datasets are not consistent. Here we would like to validate the effectiveness of our GMSD in a preclinical micro-CT application.

TABLE III
RECONSTRUCTION PSNR/SSIM/MSEs OF CIRS PHANTOM DATA USING DIFFERENT METHODS AT 60, 90, 120 AND 180 VIEWS.

| Views | FBP | U-Net | FBPConvNet | GMSD |
|---|---|---|---|---|
| 60 | 17.83/0.4980/0.01650 | 30.68/0.9563/0.00086 | 26.72/0.9334/0.00230 | **34.64/0.9698/0.00037** |
| 90 | 23.66/0.5986/0.00431 | 34.95/0.9736/0.00032 | 32.09/0.9577/0.00062 | **40.10/0.9884/0.00010** |
| 120 | 25.40/0.6878/0.00289 | 37.16/0.9794/0.00019 | 35.31/0.9732/0.00030 | **42.62/0.9923/0.00006** |
| 180 | 28.90/0.8145/0.00129 | 40.60/0.9834/0.00009 | 38.81/0.9849/0.00013 | **46.03/0.9957/0.00003** |

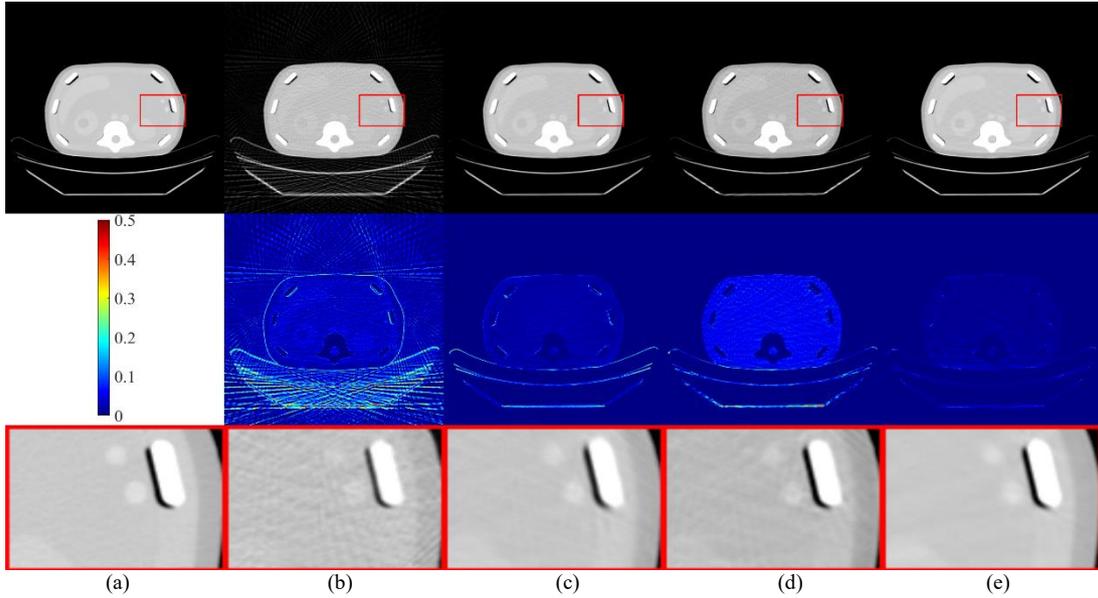

**Fig. 10.** Reconstruction images from 120 views using different methods. (a) The reference image versus the images reconstructed by (b) FBP (c) U-Net (d) FBPConvNet, and (e) GMSD. The display windows are [-650,250]. The second row is residuals between the reference images and reconstruction images.

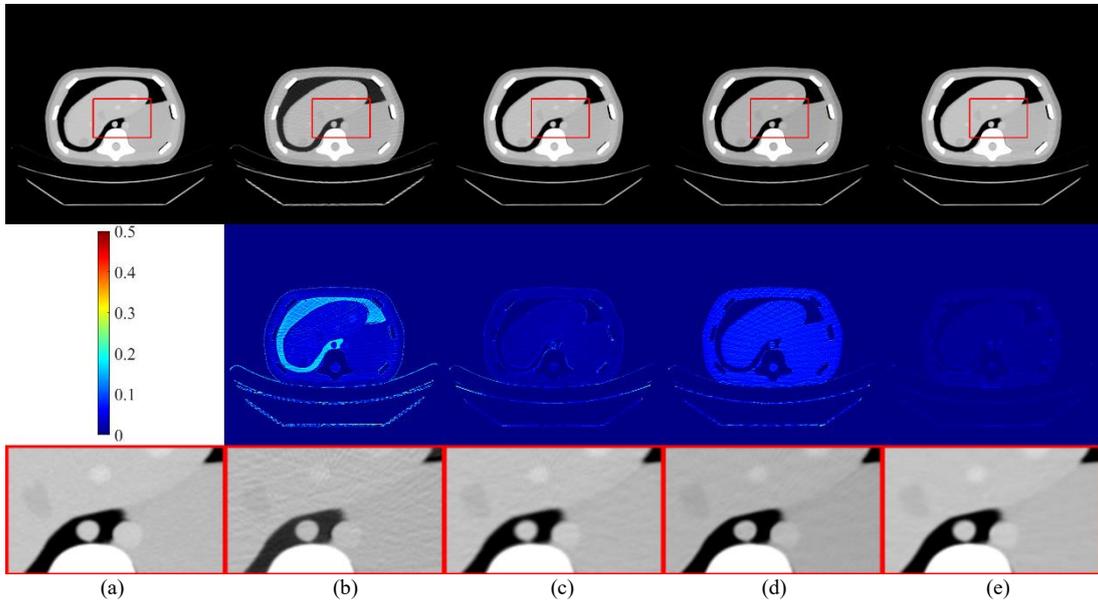

**Fig. 11.** Reconstruction images from 180 views using different methods. (a) The reference image versus the images reconstructed by (b) FBP (c) U-Net (d) FBPConvNet, and (e) GMSD. The display windows are [-650,250]. The second row is residuals between the reference images and reconstruction images.

A dead mouse is scanned by an industrial CT system with a micro-focus x-ray source and a flat-panel x-ray detector. Because the industrial CT system is designed for material non-destructive test, the x-ray source specification did not well match the requirements of preclinical mouse study. However, this mismatch does not affect the evaluation of the proposed algorithm for sparse-view CT reconstruction. The distances between the source to the detector and object are 1150 $mm$ and 950 $mm$, respectively. The detector consists of $1024\times 1024$ pixels. The size of detector bin is $0.2\times 0.2$ $mm^2$. The projection contained 500 views within the angular range $[0, 2\pi]$. The reconstructed image is a matrix of $512\times 512$ with $0.15\times 0.15$ $mm^2$ per pixel.

To test the efficiency in real data, the network in GMSD is trained on AAPM Challenge Data and test on Preclinical Mouse data. Since there are 500 views distributed within the angular range $[0, 2\pi]$, a subsampling factor is set to 5, and we obtain 100 views over the scanning range. The reconstructed results using different methods are shown in Fig. 12. The reference is reconstructed from full-view projections. It can be seen in Fig. 12 that there is still a noisy appearance in the FBP reconstruction. The extracted ROIs in Fig. 12 show that edges are blurred by FBPConvNet and HDNet. On the other hand, GMSD produced clearer edges with higher accuracy than others.

### G. Variants of Hyperparameters

The channel-copy operator maps sparse-view data to high-dimensional space, which provides richer data samples. Multi-channel can increase the number of samples and improve the accuracy of score matching. Specifically, the manifold hypothesis states that data in the real world tend to concentrate on low-dimensional manifolds embedded in an ambient space. There are two vital issues faced by score-based generative models. On one hand, when image x is restricted to a low dimensional manifold, the score is undefined, which is a gradient taken in the ambient space. On the other hand, a consistent score estimator is afforded only when the support of the data distribution is the whole space; If the data resides on a low-dimensional manifold, it will be inconsistent. In HGGDP [48], we find that by constructing and processing noise priors in high-dimensional space, it is beneficial to solve the problems of low-dimensional manifolds and low-data density regions in the generative model. Following this idea, we apply the same approach in GMSD. The reconstruction results regarding the number of input channels of the network $S_\theta(x)$ are investigated in Table IV. For 60 views, the data in 2-channel has the highest PSNR value, and SSIM value is slightly lower than 1-channel. In addition, the 2-channel obtains the best values on the 90, 120, 180 views. Considering the experimental effect, the channel number is set to be 2.

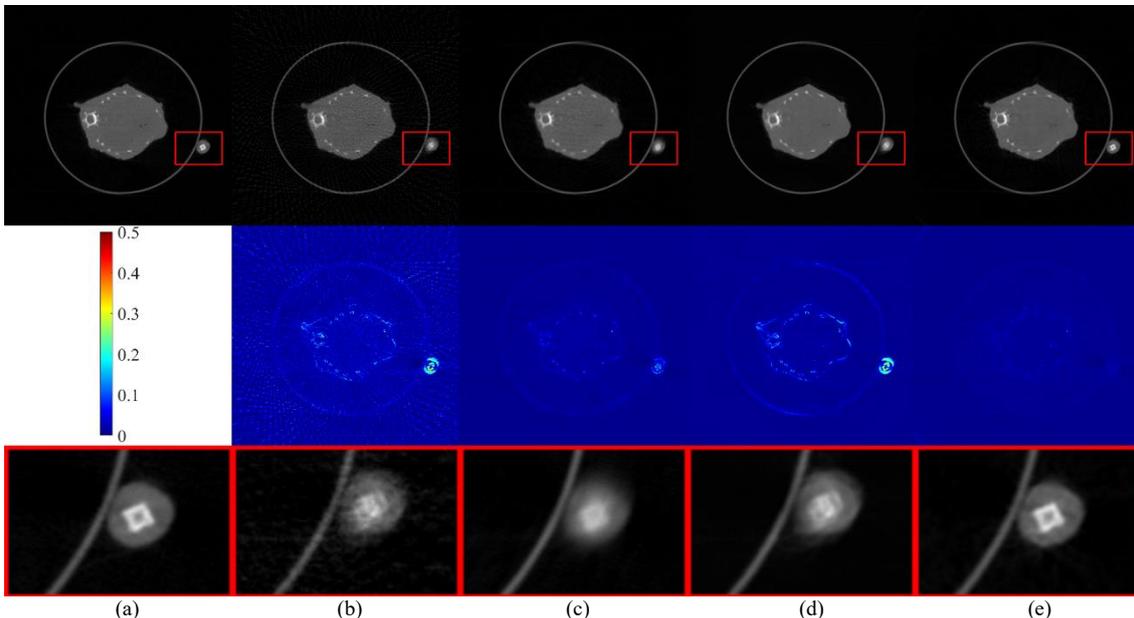

Fig. 12. Reconstruction images from 100 views using different methods. (a) The reference image versus the images reconstructed by (b) FBP (c) U-Net (d) FBPConvNet, and (e) GMSD. The second row is residuals between the reference images and reconstruction images.

TABLE IV
THE IMPACT OF CHANNEL NUMBER ON GMSD RECONSTRUCTION.

| Channel | Index | 60 | 90 | 120 | 180 |
|---|---|---|---|---|---|
| 1-ch | PSNR | 35.26 | 37.71 | 39.37 | 41.66 |
|  | SSIM | **0.9621** | 0.9753 | 0.9818 | 0.9869 |
|  | MSE | 0.00032 | 0.00019 | 0.00012 | 0.00008 |
| 2-ch | PSNR | **35.74** | **38.93** | **39.90** | **42.82** |
|  | SSIM | 0.9607 | **0.9804** | **0.9855** | **0.9909** |
|  | MSE | **0.00030** | **0.00014** | **0.00010** | **0.00006** |
| 4-ch | PSNR | 33.83 | 36.29 | 37.78 | 39.68 |
|  | SSIM | 0.9596 | 0.9718 | 0.9805 | 0.9878 |
|  | MSE | 0.00040 | 0.00025 | 0.00018 | 0.00013 |

## V. CONCLUSIONS

Although the deep learning-based CT reconstruction methods have achieved great successes in the past several years, the generalizability and robustness of trained networks is still an open problem. In this work, we presented a score-based generative model for sparse-view data tomographic reconstruction, where the PC sampling was used to generate the full-view sinogram. Specifically, we used a fully unsupervised technique to train a score-based generative model, for capturing the prior distribution of sinogram data. Then, at the iterative inference stage, the numerical SDE solver and data-consistency step was performed alternatively to achieve quantified reconstruction. More precisely, Langevin dynamics was considered as the corrector, the predictor referred to a numerical solver for the reverse-time SDE. Meanwhile, the Multi-channel strategy is adopted in the training and testing phases. Additionally, multiple noise scales were used to perturb and update the data fidelity after prediction and correction steps. The effectiveness of GMSD was verified on AAPM challenge data, CIRS phantom data and a real dataset. Experimental results illustrated that the proposed method can effectively suppress the streaking artifact and preserve image details for sparse-view CT reconstruction.

## REFERENCES

[1] M. Bakator, and D. Radosav, "Deep learning and medical diagnosis: A review of literature," *Multimodal Tech. Interact.,* vol. 2, no. 3, pp. 47 2018.
[2] D. J. Brenner and E. J. Hall, "Computed tomography—an increasing source of radiation exposure," *New Engl. J. Med.,* vol. 357, no. 22, pp. 2277-2284, 2007.
[3] E. Y. Sidky and X. Pan, "Image reconstruction in circular cone-beam computed tomography by constrained, total-variation minimization," *Phys. Med. Biol.* vol. 53, no. 17, pp. 4777-4807, 2008.
[4] H. Yu and G. Wang, "Compressed sensing based interior tomography," *Phys. Med. Biol.,* vol. 54, no. 9, pp. 2791-2805, 2009.
[5] Y. Chen, D. Gao, C. Nie, L. Luo, and W. Chen, *et al*, "Bayesian statistical reconstruction for low-dose x-ray computed tomography using an adaptive weighting nonlocal prior," *Comput. Med. Imaging Graph.* vol. 33, no. 7, pp. 495-500, 2009.
[6] J. Liu, Y. Hu, J. Yang, Y. Chen, and H. Shu, *et al*, "3D feature constrained reconstruction for low dose CT Imaging," *IEEE Trans. Circuits Syst. Video Tech*nol., vol. 28, no. 5, pp. 1232-1247. 2017.
[7] J. F. Cai, X. Jia, H. Gao, S. B. Jiang, Z. Shen, and H. Zhao, "Cine cone beam CT reconstruction using low-rank matrix factorization: algorithm and a proof-of-principle study," *IEEE Trans. Med. Imaging* vol. 33, no. 8, pp. 1581-1591, 2014.
[8] I. Y. Chun, X. Zheng, Y. Long, and J. Fessler, "Sparse-view X-ray CT reconstruction using 1 regularization with learned sparsifying transform," in *Proc. Fully Three-Dimensional Image Reconstruction in Radiology and Nuclear Medicine (Fully3D),* pp. 115-119, 2017.
[9] X. Zheng, X. Lu, S. Ravishankar, Y. Long, and J. Fessler. "Low dose CT image reconstruction with learned sparsifying transform," in *Proc. IEEE Workshop on Image, Video, Multidim. Signal Proc. (IVMSP),* pp. 1-5, 2016.
[10] E. Kang, J. Min, J. C. Ye, "A deep convolutional neural network using directional wavelets for low-dose X-ray CT reconstruction," *Med. Phys.*

vol. 44, no. 10, pp. 360-375, 2017.
[11] L. Huang, H. Jiang, S. Li, Z. Bai, J. Zhang, "Two stage residual CNN for texture denoising and structure enhancement on low dose CT image," *Comput. Methods Programs Biomed.* vol. 184, p. 105115, 2020.
[12] H. Shan, A. Padole, F. Homayounieh, U. Kruger, R. D. Khera, C. Nitiwarangkul, M. K. Kalra, G. E. Wang, "Competitive performance of a modularized deep neural network compared to commercial algorithms for low-dose CT image reconstruction," *Nat. Mach. Intell.,* vol. 1, no. 6, pp. 269-276, 2019.
[13] W. Wu, D. Hu, C. Niu, H. Yu, V. Vardhanabhuti, G. E. Wang, "DRONE: dual-domain residual-based optimization network for sparse-view CT reconstruction," *IEEE Trans. Med. Imaging,* vol. 40, no. 11, pp. 3002-3014, 2021.
[14] H. U. Chen, Y. I. Zhang, M. K. Kalra, F. Lin, Y. Chen, *et al.*, "Low-dose CT with a residual encoder-decoder convolutional neural network," *IEEE Trans. Med. Imaging,* vol. 36, no. 12, pp. 2524-2535, 2017.
[15] Y. Han, J. Yoo, and J. Ye, "Deep residual learning for compressed sensing CT reconstruction via persistent homology analysis," *arXiv preprint arXiv:1611.06391*, 2016.
[16] E. Kang, J. Min, and J. C. Ye, "A deep convolutional neural network using directional wavelets for low-dose x-ray CT reconstruction," *Med Phys,* vol. 44, no. 10, pp. e360-e375, 2017.
[17] Y. Han and J. C. Ye, "Framing u-net via deep convolutional framelets: Application to sparse-view CT," *IEEE Trans. Med. Imaging,* vol. 37, no. 6, pp. 1418-1429, 2018.
[18] K. Simonyan, A. Zisserman, "Very deep convolutional networks for large-scale image recognition," *In ICLR*, 2015.
[19] Q. Yang, P. Yan, Y. Zhang, H. Yu, Y. Shi, *et al.,* "Low-dose CT image denoising using a generative adversarial network with Wasserstein distance and perceptual loss," *IEEE Trans. Med. Imaging,* vol. 37, no. 6, pp. 1348-1357, 2018.
[20] Z. Huang, X. Liu, R. Wang, J. Chen, P. Lu, *et al.*, "Considering anatomical prior information for low-dose CT image enhancement using attribute-augmented Wasserstein generative adversarial networks," *Neurocomputing,* vol. 428, pp. 104-115, 2021.
[21] H. Lee, J. Lee, and S. Cho, "View-interpolation of sparsely sampled sinogram using convolutional neural network," *in Proc. of SPIE,* vol. 10133, p. 1013328, 2017.
[22] A. Krizhevsky and G. E. Hinton, "ImageNet classification with deep convolutional neural networks," *in NIPS*, pp. 1097-1105, 2012.
[23] D. Xu, S. Vekhande, and G. Cao, "Sinogram interpolation for sparse-view micro-CT with deep learning neural network," *in: Conference on Medical Imaging Physics of Medical Imaging,* vol. 10948, pp. 692-698, 2019.
[24] J. Lee, H. Lee, and S. Cho, "Sinogram synthesis using convolutional-neural-network for sparsely view-sampled CT," *in Medical Imaging: Image Processing*, vol. 10574, p. 105742, 2018.
[25] B. Dong, J. Li, and Z. Shen, "X-ray CT image reconstruction via wavelet frame-based regularization and radon domain inpainting," *J. Sci. Comput.,* vol. 54, no. 2-3, pp. 333-349, 2013.
[26] H. Lee, J. Lees, H. Kim, B. Cho, and S. Cho, "Deep-neural-network based sinogram synthesis for sparse-view CT image reconstruction," *IEEE Transactions on Radiation and Plasma Medical Sciences*, vol. 3, no. 2, pp. 109-119, 2018.
[27] T. Wurfl, M. Hoffmann, V. Christlein, K. Breininger, and Y. Huang, et. al., "Deep learning computed tomography: Learning projection-domain weights from image domain in limited angle problems, *IEEE Trans. Med. Imaging,* vol. 37, no. 6, pp. 1454-1463, 2018.
[28] Y. Wang, T. Yang, W. Huang, "Limited-angle computed tomography reconstruction using combined FDK-based neural network and U-Net," *in: IEEE Engineering in Medicine and Biology Society Conference Proceedings*, pp. 1572-1575, 2020.
[29] H. Chen, Y. Zhang, W. Zhang, H. Sun, P. Liao, K. He, J. Zhou, and G. Wang, "Learned experts' assessment-based reconstruction network ("LEARN") for sparse-data CT," *arXiv preprint arXiv:1707.09636,* 2017.
[30] D. Hu, J. Liu, T. Lv, Q. Zhao, and Y. Zhang, *et al.*, "Hybrid-domain neural network processing for sparse-view CT reconstruction," *IEEE Trans. Radiat. Plasma Med. Sci.* vol. 5, no. 1 pp. 88-98, 2021.
[31] Y. Liu, K. Deng, C. Sun, and H. Yang, "A lightweight structure aimed to utilize spatial correlation for sparse-view CT reconstruction," *Preprint at arXiv:2101.07613,* 2021.
[32] Q. Zhang, Z. Hu, C. Jiang, H. Zheng, Y. Ge, and D. Liang, "Artifact removal using a hybrid-domain convolutional neural network for limited-angle computed tomography imaging," *Phys. Med. Biol.,* vol. 65, no. 15, p. 155010, 2020.
[33] A. Radford, L. Metz, and S. Chintala, "Unsupervised representation learning with deep convolutional generative adversarial networks," *in ICLR*, 2016.
[34] A. Oord, S. Dieleman, H. Zen, K. Simonyan, and O. Vinyals *et al.*, "Wavenet: A generative model for raw audio," *arXiv preprint arXiv:1609.03499,* 2016.
[35] S. R. Bowman, L. Vilnis, O. Vinyals, A.M. Dai, R. Jozefowicz, and S. Bengio, "Generating sentences from a continuous space," *In Proceedings of the 20$^{th}$ SIGNLL Conference on Computational Natural Language Learning*, pp. 1021, 2016.
[36] D. P. Kingma and M. Welling, "Auto-encoding variational Bayes," *In ICLR*, 2014.
[37] K. C. Tezcan, C. F. Baumgartner, R. Luechinger, K. P. Pruessmann, and E. Konukoglu, "MR image reconstruction using deep density priors," *IEEE Trans. Med. Imag.,* vol. 38, no. 7, pp. 1633-1642, 2018.
[38] T. Taskaya-Temizel and M.C. Casey, "A comparative study of autoregressive neural network hybrids," *Neural Networks,* vol. 18, no. 5-6, pp. 781-789, 2005.
[39] M. Asim, A. Ahmed, and P. Hand, "Invertible generative models for inverse problems: mitigating representation error and dataset bias," *in ICML*, pp. 399-409, 2020.
[40] R. Salakhutdinov and G.E. Hinton, "Deep Boltzmann machines," *in Proc. Int. Conf. Artif. Intell. Statist.,* pp. 448-455, 2009.
[41] A. Graves, "Generating sequences with recurrent neural networks," *arXiv preprint arXiv:1308.0850,* 2013.
[42] E. L. Denton, S. Chintala, and R. Fergus, "Deep generative image models using a Laplacian pyramid of adversarial networks," *in Adv. Neural Inf. Process. Syst.,* pp. 1486-1494, 2015.
[43] Y. Song, S. Ermon, "Improved techniques for training score-based generative models," *in: Advances in Neural Information Processing Systems,* vol. 33, pp. 12438-12448, 2020.
[44] Y. Song, S. Ermon, "Generative modeling by estimating gradients of the data distribution," *in: Advances in Neural Information Processing Systems,* vol. 32, 2019.
[45] Y. Song, J. Sohl-Dickstein, D. P. Kingma, A. Kumar, and S. Ermon, "Score-based generative modeling through stochastic differential equations," *in ICLR*, 2021.
[46] P. Vincent, "A connection between score matching and denoising autoencoders," *Neural computation,* vol. 23, no. 7, pp. 1661-1674,
[47] G. Parisi, "Correlation functions and computer simulations," *Nuclear Physics*, vol. 180, no. 3 pp. 378-384, 1981.
[48] C. Quan, J. Zhou, Y. Zhu, Y. Chen, and S. Wang, *et al*, "Homotopic gradients of generative density priors for MR image reconstruction," *IEEE Trans. Med. Imaging*, vol. 40, no. 12, pp. 3265-3278, 2021.
[49] F. Zhang, M. Zhang, B. Qin, Y. Zhang, and Z. Xu, *et al.*, "REDAEP: Robust and enhanced denoising autoencoding prior for sparse-view CT reconstruction," *IEEE Transactions on Radiation and Plasma Medical Sciences*, vol. 5, no. 1, pp. 108-119, 2020.
[50] Z. He, Y. Zhang, Y. Guan, B. Guan, and S. Niu, *et al.*, "Iterative reconstruction for low-dose CT using deep gradient priors of generative Model," *IEEE Transactions on Radiation and Plasma Medical Sciences*. vol. 6, no. 7, pp. 741-754, 2022.
[51] Low Dose CT Grand Challenge. Accessed: Apr. 6, 2017. [Online]. Available: http://www.aapm.org/GrandChallenge/LowDoseCT/.
[52] R. L. Siddon, "Fast calculation of the exact radiological path fora three-dimensional CT array," *Medical Physics*, vol. 12, no. 2, pp.252–255, 1985.
[53] F. Jacobs, E. Sundermann, B. de Sutter, M. Christiaens, and I.Lemahieu, "A fast algorithm to calculate the exact radiologicalpath through a pixel or voxel space," *Journal of Computing andInformation Technology*, vol. 6, no. 1, pp. 89–94, 1998.
[54] J. Adler, H. Kohr, and O. Oktem, "Operator discretization library (ODL)," Software available from https://github.com/odlgroup/odl, vol. 5, 2017.
[55] A. Beck, and M. Teboulle, "Fast gradient-based algorithms for constrained total variation image denoising and deblurring problems," *IEEE Trans. Image Process.*, vol. 18, no. 11, pp. 2419-2434, 2009.
[56] Y. Censor, and T. Elfving, "Block-iterative algorithms with diagonally scaled oblique projections for the linear feasibility problem," *SIAM J. Matrix Anal. Appl.*, vol. 24, no. 1, pp. 40-58, 2002.
[57] H. Lee, J. Lee, H. Kim, B. Cho, and S. Cho, "Deep-neural-network-based sinogram synthesis for sparse-view CT image reconstruction," *IEEE Transactions on Radiation and Plasma Medical Sciences,* vol. 3, no. 2, pp. 109-119, 2019.
[58] J. Kyong Hwan, M. T. McCann, E. Froustey, and M. Unser, "Deep convolutional neural network for inverse problems in imaging," *IEEE Trans. Image Process.* vol. 26, no. 9, pp. 4509-4522, 2017.